# How visual discomfort changes with horizontal viewing angle on stereoscopic display

Yaohua Xie[a], Danli Wang*[a], Fang Sun[b]

[a]Institute of Software, Chinese Academy of Sciences, Beijing, China

[b]School of Computer and Information Technology, Liaoning Normal University, Dalian, China

*Tel.: +86-10-139-1098-0567; E-mail: danliwang2009@gmail.com, Postal address: 4# South Fourth Street, Zhong Guan Cun, Beijing, 100190, P.R. China.

This work was supported by the Major State Basic Research Development Program of China under Grant No.2013CB328805, the National Natural Science Foundation of China under Grant No.61272325 and No.60970090.

**Abstract:** When viewing stereoscopic displays, people may not always be able to stay exactly in front of the display. It is known that viewing stereoscopic display from different vertical angles lead to different visual discomfort. However, the effects of horizontal viewing angle on stereoscopic visual discomfort have been rarely investigated, especially for household stereoscopic displays. In this study, subjects were required to view a stereoscopic display from various horizontal viewing angles, and assessed their visual discomfort during viewing. The visual stimuli have various amount of disparities: positive disparity, negative disparity or zero disparity. Results showed that the visual discomfort changes with horizontal viewing angle, and greater angles generally lead to more serious visual discomfort. Furthermore, the relationship between visual discomfort and horizontal viewing angle can be approximately expressed by a quadratic function.

# Introduction

Nowadays, there are all kinds of screens in our daily life or work, such as monitors, televisions, cinemas, and so on. In recent year, stereoscopic displays are also becoming more and more popular. When viewing a screen, many people may prefer looking from a viewing angle of zero degrees, i.e., from exactly the front of the screen. However, this need cannot be met in many cases. For example, Paul Hands et al. (2015) have analyzed the situation of both watching television at home and seeing movie in a cinema. As shown by the result, the horizontal viewing angle can be larger than 20 degrees when watching television. Furthermore, the horizontal viewing angle can be larger than 60 degrees when seeing movie in a cinema.

Many researchers have devoted into the topics relevant to this phenomenon, e.g., the effect of oblique viewing on object perception. Many artists or scientists have showed interest in the reason of perceptual invariance across viewing position with pictures (Vinci, 1970; Cutting, 1987; Zorin, 1995). Some existing studies have shown that incorrect viewing distance can lead to distortions in perceived depth and shape (Held, 2008; Woods, 1993). Some researchers have considered perceptual distortions in stereoscopic 3-D due to oblique viewing (Banks, 2009; Bereby-Meyer, 1999; Perkins, 1973). Yoella Bereby-Meyer et al. (1999) investigated the distortions of perception when artificial stereoscopic displays were seen from an inappropriate distance and/or orientation. The results indicated that people can compensate partially for distortions in stereopsis when the relevant information is available. By changing the projection angle and the position of a stimulus in the picture, Dhanraj Vishwanath et al. (2005) found that invariance is achieved through an estimate of local surface orientation, not from geometric information in the picture. Martin S. Banks et al. (2005) studied the perceived ovoids and planes in pictures while changing viewing angle and the angle by which the pictures were projected. They also changed the viewer's information about the orientation of the picture surface. According to the results, they concluded that the subjects adopted an estimate of local surface orientation and not prior information for object shape nor geometric information in the picture. In their article, Martin S. Banks et al. (2009) reviewed existing research on the ability of a viewer to perceive the 3-D layout presented on a stereo display. Paul Hands et al. (2013) found that stereoscopic content may look warped when viewed from oblique angles, but the effect is small and non-stereo cues can help to restore the effect. Paul Hands (2015) also found that, at least for object distortion, oblique viewing is unlikely to cause substantially a more serious problem for S3D content than it already is for 2-D.

Beside perception of shape, viewing obliquely may also have effects on other aspects of observer experience, e.g., visual discomfort or visual fatigue. In recent decades, the problem of visual fatigue or visual discomfort has been extensively investigated. It is generally believed that accommodation-vergence conflict is an important source of visual fatigue and discomfort (Yano, 2004; Emoto, 2005; Hoffman, 2008; Shibata, 2005; Wann, 2002). In some studies, researchers intend to distinguish the term visual discomfort from the term visual fatigue; although they are often interchangeable. Lambooij et al. (2009) define visual fatigue as a decrease in performance of the human vision system, which can be objectively measured, while define visual discomfort as the subjective counterpart of visual fatigue. Urvoy et al. (2013) think that visual discomfort is perceived instantaneously, while visual fatigue is induced after a given duration of effort. Our study is mainly focused on visual discomfort.

Visual discomfort can be influenced by many factors. When the position of the observer changes, the direct factors are the viewing distance and the viewing angle. Takashi Shibata et al. (2011) used a novel volumetric display to investigate how viewing distance and the sign of the vergence-accommodation conflict affect visual discomfort and fatigue. They observed more discomfort and fatigue with a given vergence-accommodation conflict at the longer distances, and more serious symptoms with uncrossed disparities at long distances and with crossed disparities at short distances. Celestine A. Ntuen et al. (2009) conducted an experiment which was a 2 x 2 within subjects design comparing performance between 2-D and 3-D display modes at a viewing angle of 0 degrees and 20 degrees. Visual fatigue was also assessed. The results show that the visual fatigue with autostereoscopic display is similar to other TFT-LCD displays. The viewing angle varied vertically in their study. In our study, we are curious about the effects of horizontal viewing angle on visual discomfort. Such situation is very common when more than one person are viewing the same screen together, and it is also meaningful for choosing the proper viewing position for solo person.

## Method

### Apparatus

The experiment was performed in a room with similar lighting condition to normal office. The temperature was about 26 degrees Celsius. All the apparatuses were placed in this room. Stimuli were presented at a distance of 87 cm, on a LG stereoscopic display with a size of 0.51m x 0.29m and screen resolution of 1920 x 1080. The distance from the eyes of the subject to the center of the screen was about three times of the screen height, which is recommended by ITU-R BT.2021-1 (Union 2015). The display can rotate horizontally while keeping the distance from its center to the eyes fixed. Five angles were adopted in the experiment: -45, -30, 0, 30, and 45 degrees, respectively. A desktop PC was used to present stimuli and record the response of the subjects. A desk and a chair were set properly so that subjects could easily keep correct position relative to the screen. The definition of horizontal viewing angle is shown by Fig. 1.

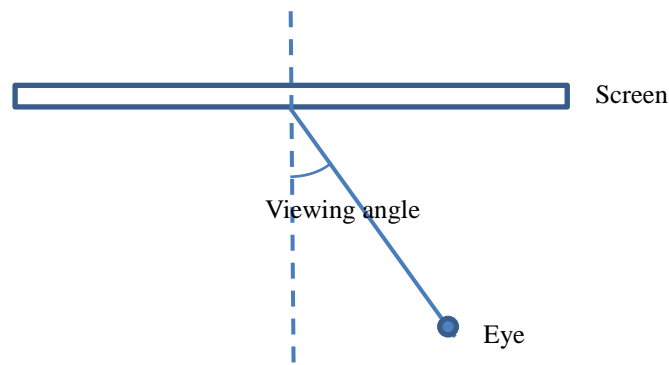

**Figure 1. The definition of horizontal viewing angle.**

## Subjects

Sixteen naïve subjects (10 male and 6 female) took part in the experiment, and valid data were collected for fourteen subjects (9 male and 5 female) finally. The fourteen subjects were ranged in age from 22 to 43 years; the mean of age was 29.4 years. All participants were paid for their participation measured by time period, and they were all with normal or correct-to-normal vision. The entire study was approved in accordance with the Declaration of Helsinki.

## Stimuli

The visual stimuli used in the experiment were photos with various disparities. Subjects were required to observe the main target of every photo which located in the center. Some of the photos have main objects with positive disparities, some with negative disparities, and the others with zero disparities. In the experiment, these photos were observed by the subjects from various horizontal viewing angle, and the presentation order of these photos was determined randomly.

## Procedures

In the beginning of the experiment, subjects must have a test to determine whether their stereo vision met our requirement. They could take part in the rest of the experimental procedure if their stereo acuities were better than 200". After the test, the procedure was as shown by Fig. 2.

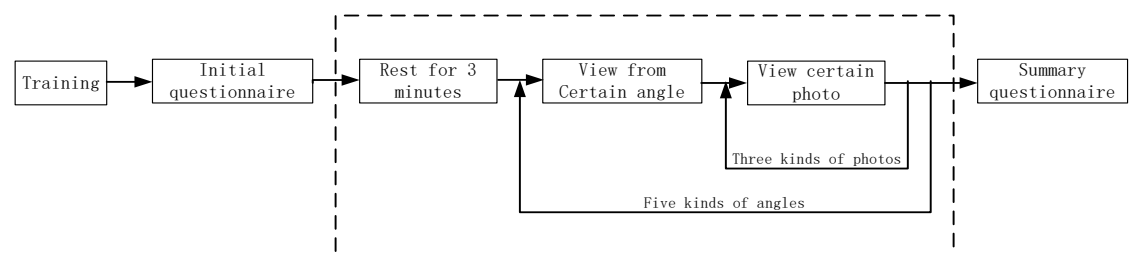

**Figure 2. The procedure of the experiment.**

In the beginning, the subject was shown a group of photos, and he/she was required to practice rating his/her visual discomfort during viewing. This step was designed for the subject to understand the rules of the experiment, and also get familiar with the mapping between the visual discomfort and the rating scores. The photos used in this step was not same as those used in the rest of the procedure. After training, the subject was required to answer an initial questionnaire. The experimenter explained every question in the questionnaire and instructed the subject to answer it step by step. The initial questionnaire was used to collect the personal information of the subject, and also recorded the current visual discomfort of the subject including the detailed symptoms. The current scores of visual discomfort was then used as the baseline in the analysis of the results.

After three minutes of rest, the formal stimuli were then presented to the subjects. Every subject needed to observe the stimuli from different horizontal viewing angles, and three photos were presented for every angle. These photos were comprised of three different kinds, which had positive disparity, negative disparity or zero disparity in the target area, respectively. Each photo was presented for about one minute. For each subject, the order of the viewing angle was determined randomly. For each viewing angle, the order of the presented photos was also determined randomly. Every time when a certain photo was viewed, the subject was required to rate his/her visual discomfort at that time. Then the subject can take a rest of three minutes again. After that, he/she was required to rate the visual discomfort once more. The above steps were repeated until all the viewing angle had been tested.

After all the stimuli were viewed, the subject was required to evaluate the visual discomfort including the detailed symptoms through the summary questionnaire. The scores of visual discomfort were defined similar to the SSCQE regulation in ITU-R BT.2021-1 (Union 2015) which could be one of the values listed in Table 1.

| 1 | 2 | 3 | 4 | 5 |
|---|---|---|---|---|
| Very comfortable | Comfortable | Slightly uncomfortable | Uncomfortable | Very uncomfortable |

**Table 1. The possible values of visual discomfort.**

## Results

In our experiment, there were photos with three different kinds of disparity. For each kind of disparity, rating scores of visual discomfort were collected for all the aforementioned fourteen subjects. These scores represented the amounts of visual discomfort before viewing, after viewing a certain photo or after taking a rest.

At the beginning of the analysis, we took two steps of pre-process to make the data more meaningful. On the one hand, the original scores may not represented the visual discomfort caused by a single photo as the visual discomfort may accumulate during time. So we subtracted a starting score from each score for any single photo. The starting scores were the scores for the moments when the subject had just finished rest. On the other hand, the scores of different subjects may have different valid range. Therefore, we normalized the scores for all the subjects so that all the scores were within [0, 1]. After pre-process, we analyzed the scores for three different disparities respectively, and the result is shown by Fig. 3.

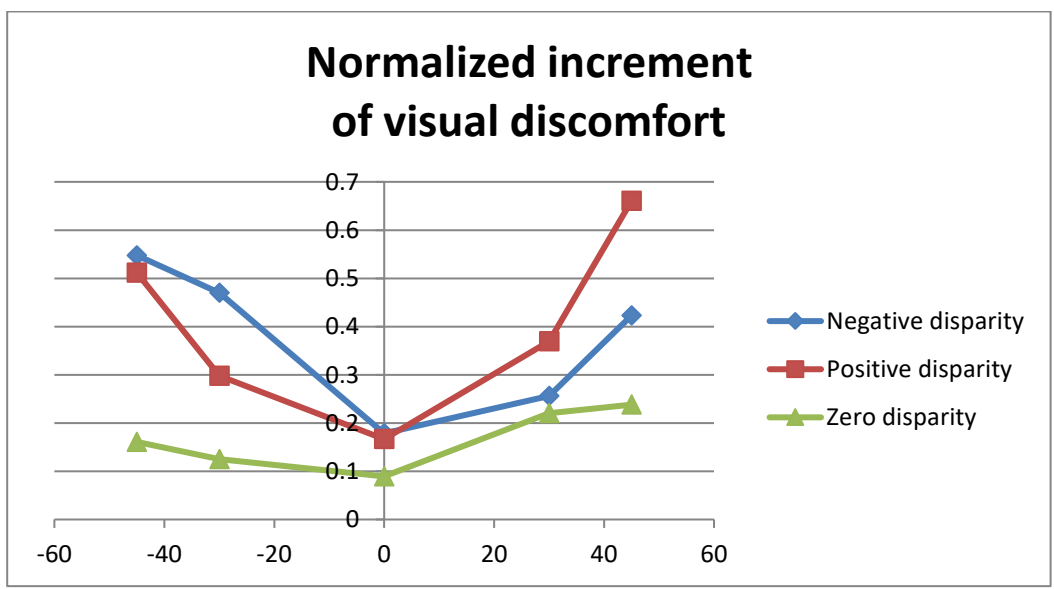

**Figure 3. Normalized increment of visual discomfort for three different disparities.**

Fig. 3 shows the effects of horizontal viewing angle on the corresponding visual discomfort. In all the three curves, the scores of the visual discomfort varied with horizontal viewing angle. In general, the visual discomfort corresponding to zero disparity was always the smaller. When the horizontal viewing angle became larger, the visual discomfort also showed a tendency of increase. The maximal value of visual discomfort appeared in the horizontal viewing angle of 45 or -45 degrees. These characteristics appeared in all the three curves.

Furthermore, we calculated the average scores of all the three disparities because we were more interested in the more general conclusion. The average scores of the normalized increment of visual discomfort are shown by Fig. 4.

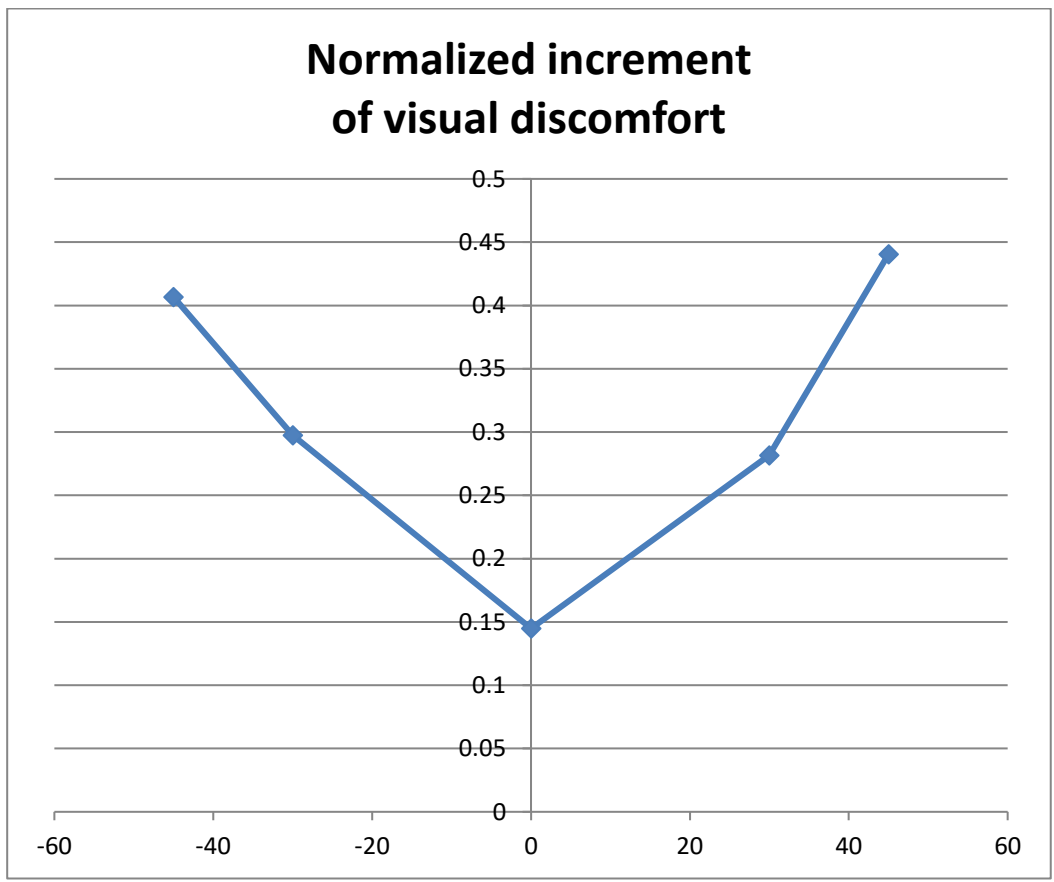

**Figure 4. Normalized increment of visual discomfort for all the photos.**

As shown by Fig. 4, the characteristic of the normalized increment of visual discomfort for all the photos is very similar to the results for the photos corresponding to each kind of disparity. It can be seen that the average scores also showed a similar tendency to the above three curves. But it was more symmetrical and more regular. The lowest point lied in the horizontal viewing angle of zero. Similar score of visual discomfort appeared in the points of -30 and 30 degrees. The scores in the points of -45 and 45 degrees were also near to each other. The tendency of the curve indicated that the score of visual discomfort increase gradually with horizontal viewing angle.

We further did ANOVA for the scores, and the result showed that there was a significant difference between the score of horizontal viewing angle zero and horizontal viewing angle 45 ($p = 0.048$). And there was nearly a significant difference between that of horizontal viewing angle zero and horizontal viewing angle -45 (p = 0.074).

As can be seen from the figure, the general curve has a profile similar to a quadratic curve. Therefore, we further performed an interpolation for the average score. The result of ANOVA of the quadratic model is shown by Table 2, and the resulting coefficients are shown in Table 3.

**ANOVA**

|  | Sum of Squares | df | Mean Square | F | Sig. |
| --- | --- | --- | --- | --- | --- |
| Regression | .054 | 2 | .027 | 53.919 | .018 |
| Residual | .001 | 2 | .000 |  |  |
| Total | .055 | 4 |  |  |  |

**Table 2. The result of ANOVA of the quadratic model.**

As can be seen from Table 2, the result of the quadratic model is significant (p = 0.018), which indicates the effectiveness of the model.

According to Table 3, the equation of the quadratic model is as follows:

$$d = 1.58 \times 10^{-1} + 1.78 \times 10^{-4} * a + 1.34 \times 10^{-4} * a^2 \tag{1}$$

Where, $d$ represents the visual discomfort, and $a$ represents the horizontal viewing angle. Beside the coefficients, the significances of the coefficients are also shown in Table 3. These values demonstrate the effectiveness of the model.

**Coefficients**

|  | Unstandardized Coefficients | | Standardized Coefficients | t | Sig. |
| --- | --- | --- | --- | --- | --- |
|  | B | Std. Error | Beta |  |  |
| Horizontal viewing angle | .000 | .000 | .058 | .611 | .603 |
| Horizontal viewing angle ** 2 | .000 | .000 | .989 | 10.367 | .009 |
| （Constant） | .158 | .018 |  | 8.712 | .013 |

**Table 3. The coefficients of the quadratic model.**

The profile of the quadratic model is shown in Fig. 5. It can be seen that the proposed model fits the experimental data well.

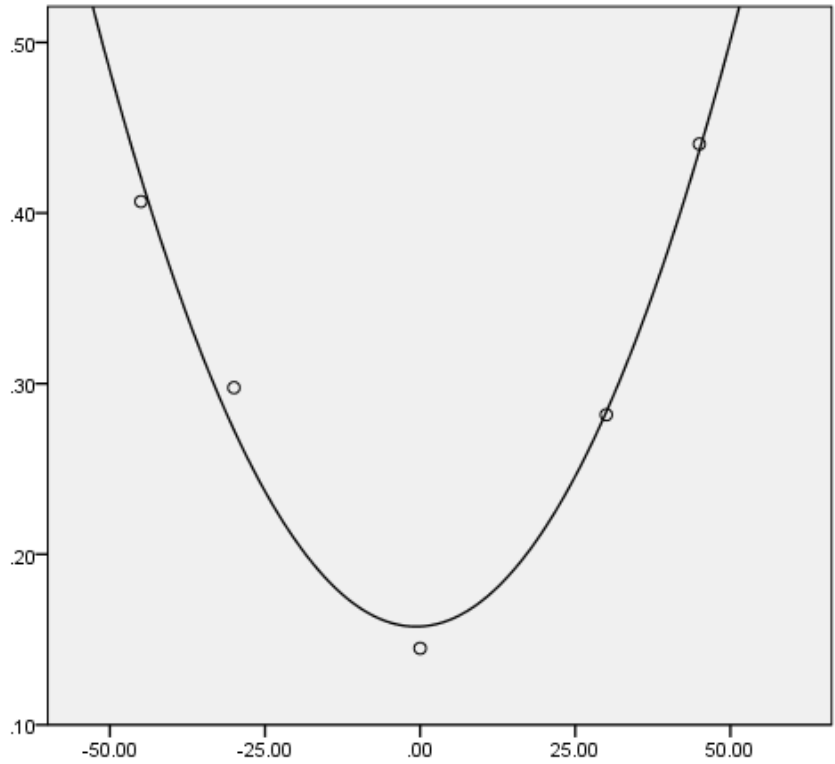

**Figure 5. The profile of the quadratic model.**

## Discussion

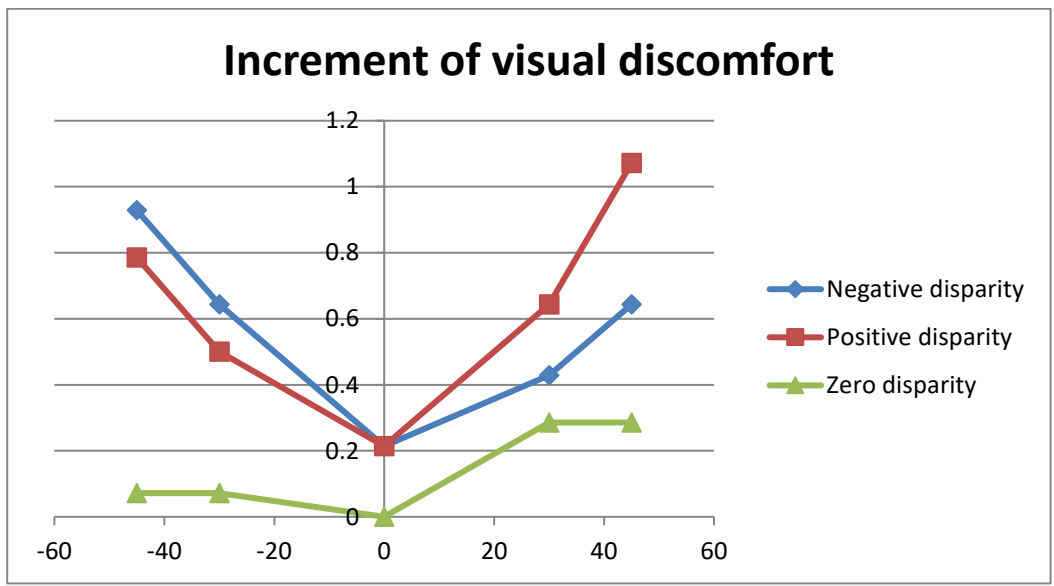

**Figure 6. Increment (without normalization) of visual discomfort for three different disparities.**

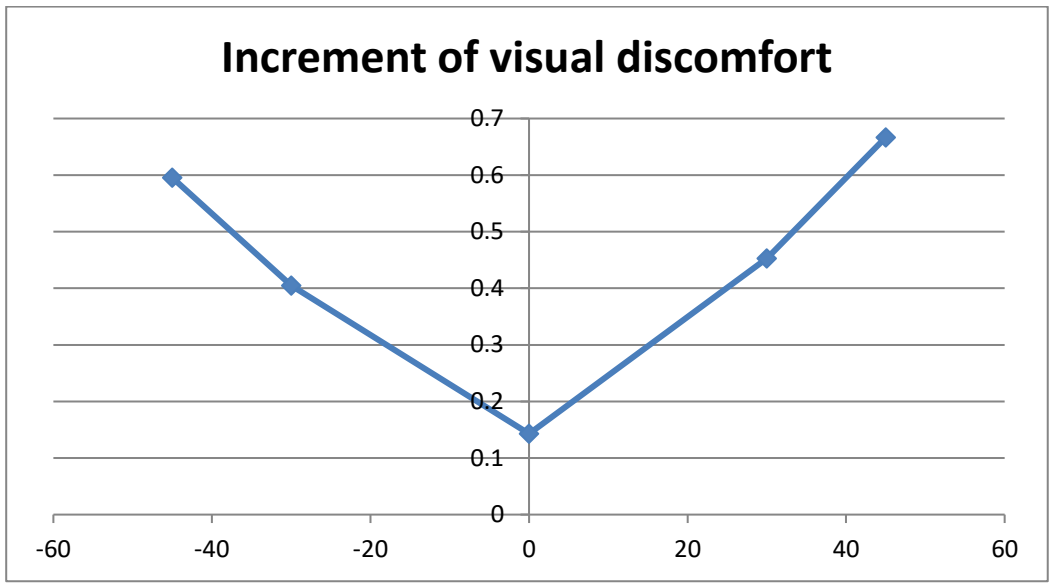

**Figure 7. Increment (without normalization) of visual discomfort for all the photos.**

As shown by the above results, the visual discomfort varies with horizontal viewing angle when viewing stereoscopic display. In general, the lowest visual discomfort appears at the horizontal viewing angle of zero. When the horizontal viewing angle increases, the visual discomfort also increases gradually. In our experiment, the visual discomfort at the horizontal viewing angle of zero has significant or nearly significant differences with that at the horizontal viewing angle of 45 or -45 degrees. Furthermore, the profile of the visual discomfort shows a similar shape of quadratic curve. Therefore, we built a regression model based on quadratic function for the experimental data. Although the curve looks nearly symmetrical, there is still a tiny difference between the scores corresponding to the horizontal viewing angle of -45 degrees and 45 degrees, and also between those corresponding to the horizontal viewing angle of -30 degrees and 30 degrees. Such differences may be partly because of the slight difference between the light distributions in the different part of the room. But there may also be other reasons which need further investigation.

In our analysis, we normalized the data so as to make them more comparable. However, the profile of the visual discomfort is actually rather stable. Even without normalization, similar pattern can also be observed. The result without normalization is shown by Fig. 6 and Fig. 7.

The change of the visual discomfort may be caused by several reasons. First of all, the retinal image may be distorted to some extent when a real world scene is projected in to an eye especially when viewing obliquely. The larger the horizontal viewing angle, the more serious the distortion. It may lead to extra burden on the vision system, and then cause visual discomfort.

Secondly, the distortions of the retinal images in the left eye and the right eye are usually different from each other. When the display is viewed from a larger horizontal viewing angle, the difference of the distortion will be larger than when viewed from smaller horizontal viewing angles. As a result, the retinal images of the two eyes are also harder to be fused to produce a stereoscopic image. In order to achieve the fusion, the vision system may need extra effort and thus may cause extra visual discomfort.

Thirdly, when the display is viewed obliquely, the illumination of the screen may vary with horizontal viewing angle. It has been known that illumination has effect on visual comfort (Zheng Yan et al., 2008). Therefore, this may contribute to the change of visual discomfort.

## Conclusions

The effects of horizontal viewing angle on the visual discomfort of stereoscopic display were investigated in this study. The subjects were required to view photos from different horizontal viewing angle and assess the visual discomfort during viewing.

The results demonstrated that horizontal viewing angle have significant effects on the visual discomfort of stereoscopic display. When viewing exactly from the front, the visual discomfort is the least serious. When the horizontal viewing angle increases, the visual discomfort also increases accordingly. As the profile of the visual discomfort shows similar shape to quadratic curve, we further build a regression model based on the quadratic function.

## Acknowledgments

This work was supported by the Major State Basic Research Development Program of China under Grant No.2013CB328805, the National Natural Science Foundation of China under Grant No.61272325 and No.60970090.